# Physicists Create Closed Time-like Curves

Jeremy D. Schnittman (NASA GSFC)

Two centuries after the initial formulation of general relativity, and exactly one hundred years after the remarkable first detection of gravitational waves, Albert Einstein is still rocking the physics world. Physicists today reported the experimental confirmation of what may be relativity's most outrageous prediction: time travel. Appropriately enough, the breakthrough was announced at the Einstein Institute of Applied Relativity (EIAR) in Jerusalem. The Institute's co-directors, Drs. Mahmoud Habibi and Shoshana Haber, were joined by Dr. Kenji Nakama of the International Space Agency (ISA), a major partner in the collaboration.

"Building on the past, we have literally gained a glimpse into the future," declared a jubilant Dr. Habibi. "It is with tremendous excitement that we report our successful demonstration of a closed time-like curve, better known as a time machine. But before we get too excited about going off to rewrite history and save humanity, a few important caveats are in order."

"First, this demonstration is one of extremely small scale, both in mass and time. We have successfully sent subatomic particles back in time, and only by fractions of a second. It is not clear that macroscopic objects like people will ever be able to travel back in time. Second, this technique can only ever be used to send particles to some time *after* the experiment was constructed. In other words, while we cannot change the past, the future may well be able to affect the present."

Dr. Haber then went on to describe some of the many important advances in theoretical physics and technology that combined to make a genuine time machine a reality.

## Gödel-Tipler metric

In 1949, Kurt Gödel derived a most curious solution to the Einstein field equations, representing a rotating universe. While not consistent with observations of our own universe, Gödel's solution did not violate any of the laws of relativity. And yet it still allowed for the existence of *closed time-like curves*. "Time-like curves" are any paths through space that are permitted for massive particles like electrons, protons, or spaceships. "Closed" time-like curves are any paths that end up back where and *when* they started, hence closed curves in the 4-dimensional world of space and time.

Gödel's hypothetical universe received increasing attention in the latter part of the 20th century as general relativity blossomed into a field of active research. In 1974, a graduate student named Frank Tipler discovered a similar solution to Einstein's equations, based on an infinitely long rotating cylinder. Unfortunately, interest in Tipler's metric waned around the turn of the 21st century when Stephen Hawking proved that any *finite* version required exotic matter with negative density to work.

This problem was overcome---at least in theory---in the mid-21st century when Peng Jiahui showed that the Gödel and Tipler solutions could be combined with concentric, counter-rotating cylinders made of *ordinary* matter, with the region between the two cylinders allowing for the existence of closed time-like curves. The only problem was that both cylinders need to rotate at the speed of light, another practical challenge.

## Helical laser cavity

In 2061, when working on the first gigahertz gravitational-wave detector, a team at Corning Optical invented the helical laser cavity (HLC), which itself was based on many of the important technology developments that made the early gravitational-wave experiments like LIGO possible. One of the important features of the HLC is that photons propagating in opposite directions are circularly polarized, like two intertwined spiral staircases: one for ascending and one for descending, yet never crossing.

Over the subsequent fifty years, much greater laser power has been achieved, to the point where the effective density of the photon beam far exceeds that of any known material. This means that HLCs can actually warp the spacetime around them. Yet the angular momentum of the counter-polarized beams exactly cancels out, making them ineffective as Tipler cylinders.

That is, until about twenty years ago, when the Corning team and a group at the National Institute for Standards and Technology (NIST) independently discovered a way to physically separate the two beams, essentially constructing a Gödel-Tipler spacetime with two concentric, counter-rotating cylinders made of pure laser energy, naturally "spinning" at the speed of light (see Figure 1).

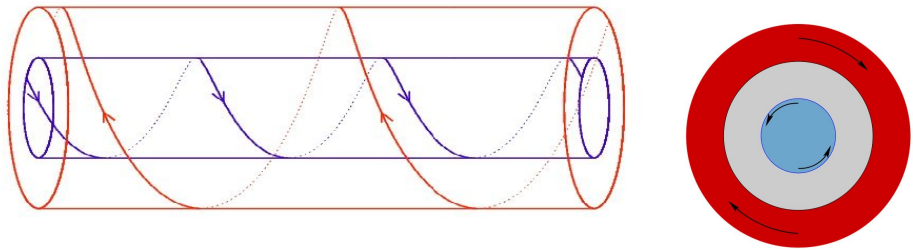

Figure 1: A split helical laser cavity, viewed from the side (left) and end-on (right). The light is circularly polarized, so the cavity behaves exactly like two concentric cylinders rotating at the speed of light. The grey region between the cylinders is described by the Gödel-Tipler metric.

## Dark matter trapping

While the Corning and NIST teams likely created regions of closed time-like curves with their split cavities, at the time there was no way to actually probe the region between the two light cylinders. Due to the extreme laser power making up the outer cavity, any external probe or particle injected into the intermediate region would immediately destroy the system. It was understood that dark matter particles would be ideal candidates for this task, as they could pass effortlessly through the outer cylinder.

At the time, this idea was absurd, since the very properties that make these weakly interacting particles attractive for penetrating the photon barrier also make them nearly impossible to control or detect. Yet ever since the demonstration of resonant terahertz transmission of gravitational waves by the WEBER team at the University of Maryland in 2093, not only has the world's communication network been completely revolutionized, but many of the technological applications once reserved for electromagnetic radiation have been extended to the gravitational-wave regime.

In particular, gravitational-wave resonators can now be used to trap and manipulate dark matter particles in much the way that ultracold atoms were first trapped with lasers over a hundred years ago (in the process confirming yet another prediction of Einstein!). This allows experimenters to craft beams of dark matter particles with finely-tuned pulse profiles and inject them into the HLC. However, when the Corning lab attempted doing just this, they found that the laser field---and thus the time machine---collapsed every time.

Eventually they realized that this collapse was a direct result of the Hawking-Thorne theorem, better known as the "grandfather paradox." Still remarkably productive at the age of 100, in 2040 Kip Thorne was finally able to prove the chronology protection conjecture first put forth nearly fifty years before by Stephen Hawking. Simply put, the global structure of space-time must be causally unique: every event has a single past and a single future. Any attempt by man or nature to circumvent this structure will inevitably fail.

## Quantum entanglement

For the first 180 years of its life, general relativity insisted on being a classical theory, stubbornly refusing a forced marriage with quantum mechanics. After generations of fruitless effort with string theory, brane theory, and loop quantum gravity, in 2095, at the age of 20, the brilliant young prodigy Claudia de Luna Amigo turned the world on its head with a completely new approach based on entanglement.

What Einstein famously called "spooky action at a distance" is an apparent paradox between the finite speed of light in relativity, and the seemingly instantaneous collapse of a quantum system: when two particles are prepared together, observing the properties of one particle will immediately determine those of the other, even if they have been separated by many kilometers. De Luna Amigo resolved this paradox with another exotic prediction of general relativity: wormholes. Just like Feynman's path integral description of quantum mechanics, Amigo's theory of quantum gravity is based on every point in space-time being connected to every other point with an infinite number of wormholes. In quantum gravity, the fabric of spacetime is constructed to minimize the non-causal chronology of the universe.

Haber and Habibi's great breakthrough came only a few years later in 2098 when they proposed to use quantum entanglement to conserve the total amount of causality in the Gödel-Tipler space-time. By using two entangled dark matter particles, one injected in a clockwise orientation, and one in a counter-clockwise orientation, the system remains stable against chronology collapse. In effect, one particle goes back in time, and one jumps forward in time exactly the same amount. This has the fortuitous additional benefit of conserving energy in the four-dimensional structure of space-time: while a particle traveling back in time effectively adds mass and energy to the past, it can safely do so by "borrowing" from the future by sending its partner particle an equal distance forward in time, so the global balance sheet always evens out.

## Gravitational isolation

One crucial problem remained. The injected particles spend so little time inside the cavity that the net time travel was insignificant relative to the total travel time from injection to detection, making it impossible to definitively prove that the particle did indeed traverse a closed time-like curve. The trick was to inject the particle with such a small velocity that it could actually orbit inside the HLC cylinders, deflected by the small gravitational field of the intense laser beams. And to do this, all other gravitational forces must be perfectly cancelled.

Yet again, gravitational-wave technology was the answer. LISA, the Laser Interferometer Space Antenna, was the first gravitational-wave detector to fly in space in 2030. In order to shield the experiment from outside disturbances, each spacecraft was designed to isolate a pair of test masses that would then be allowed to follow perfect geodesic trajectories, feeling only gravitational forces.

Over the past ten years, ISA has spent billions of €urasios on the construction of "L5," a massive science laboratory located 150 million km away at the Earth-Sun Lagrange point, designed on isolation principles similar to those that made LISA possible. At the beginning of last year, the EIAR experiment was launched to L5 and commissioning began. In another poetic coincidence, full scientific operations started in November 2115, exactly 200 years after Einstein's publication of the gravitational field equations.

As shown in Figure 2, pairs of entangled dark matter particles are injected in short pulses once per second. They are deflected by the small gravitational field in the gap between the two cylinders, and then exit the device, to be detected by the array of gravitational-wave resonators arranged around the chamber. Particles that spend longer inside the gap are deflected more, and travel further in time.

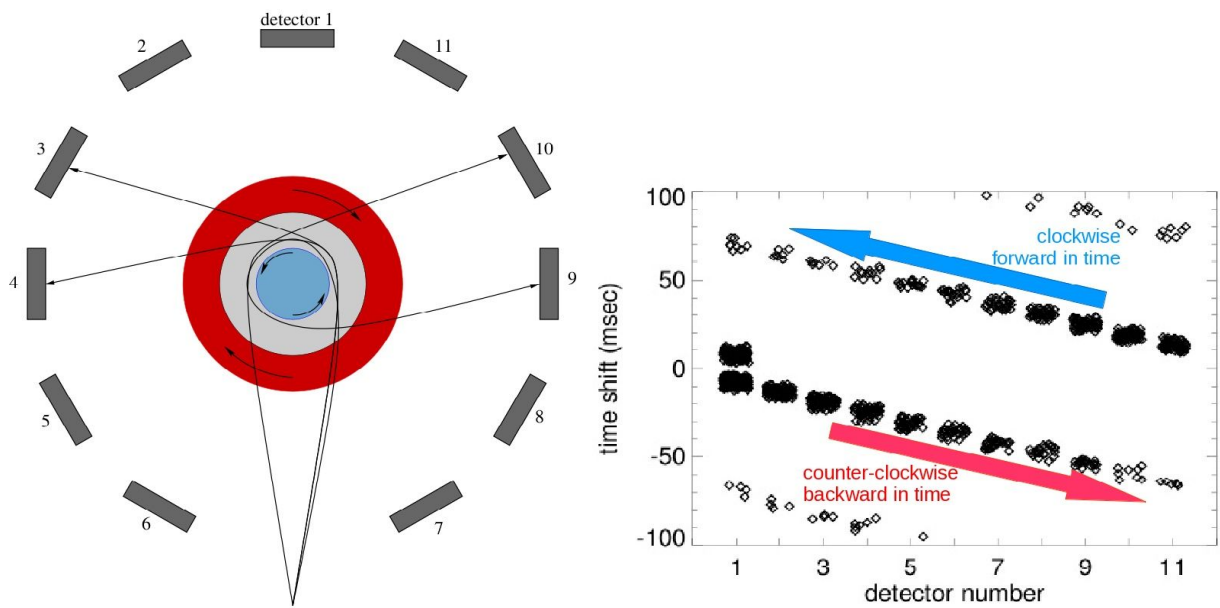

Figure 2: The EIAR experiment at the ISA L5 laboratory (left). Particles are injected from below into the HLC interaction region (grey), where they are deflected by the gravitational field of the central laser cavity (blue). Detectors (dark grey rectangles) are arranged around the experiment to record the time at which each particle escapes. Initial data is shown in the right-hand panel, clearly demonstrating negative time shifts.

While the experimental data clearly shows that time travel is indeed possible, there are still some major obstacles to overcome before we can make a fortune by betting on next year's Super Bowl. In particular, the trajectories inside the HLC are highly unstable, so it is very difficult to aim the injections just right so that the particles spent more than a few hundred milliseconds in the cavity (note how few events in Fig. 2b complete an entire orbit).

Yet Dr. Haber expressed unreserved optimism. "It's definitely too early to say for sure, but I can report that we have already seen some very anomalous events, with time shifts far beyond the expected distribution. I admit this sounds a little crazy, but we think we might be seeing evidence of injections from the future. For this reason, the collaboration has unanimously agreed to keep the experiment running continuously over the coming year, and are scheduling periods of 'down time' where the HLC is held in place, but no particles are injected. Then our future selves can work out some of the stability problems, and simply send the answers back to the present."

Dr. Habibi concluded the press conference with the reflection, "A hundred years ago, when LIGO first detected gravitational waves, Kip Thorne was extremely cautious when asked about their practical applications. I am sure Kip would have been overwhelmed to see the role they play in our lives today, and especially thrilled to see their role in time travel. I trust there will be no such questions about practical applications now. And best of all, we can guarantee that no grandfathers were harmed in the course of this work!"

*I gratefully acknowledge the valuable comments and support of Zachary Shrier and Shane Larson.*